\documentclass{PoS}
\usepackage{graphicx, subfigure} 
\usepackage{cite}

\ShortTitle{Nf=12 QCD}

\title{Chiral symmetry of QCD with twelve light flavors}

\author{A. Deuzeman \footnote{Present address: Institute for Theoretical Physics, Ch-3012 Bern, Switzerland}\\
 Centre   for   Theoretical   Physics,
University  of  Groningen,  9747  AG, Netherlands\\
        E-mail: \email{a.deuzeman@rug.nl}}

\author{\speaker{M. P. Lombardo}\\
 INFN-Laboratori Nazionali di Frascati, I-00044, Frascati (RM),
Italy \\
        E-mail: \email{lombardo@lnf.infn.it}}

\author{E. Pallante\\Centre   for   Theoretical   Physics,
University  of  Groningen,  9747  AG, Netherlands\\
        E-mail: \email{e.pallante@rug.nl}}

\abstract{ We study QCD with twelve light flavors at intermediate values of the bare lattice coupling. We contrast the results for the order parameter with different theoretical models motivated by the physics of the Goldstone phase and of the symmetric phase, and we perform a model independent analysis of  the meson spectrum inspired by  universal properties of chiral symmetry. 
Our analysis favors chiral symmetry restoration.}

\FullConference{The XXVIII International Symposium on Lattice Field Theory, Lattice2010\\
		June 14-19, 2010\\
		Villasimius, Italy}

\begin{document}

\section{Introduction}
In this note --which is part of an extended publication 
\cite{Deuzeman:2009mh}
-- we discuss the realization of chiral symmetry in QCD with twelve  flavors.  This is an ongoing effort which aims at clarifying the phase diagram of QCD with fundamental fermions - effort  motivated either by theoretical and phenomenological considerations, nicely reviewed in e.g. \cite{Sannino:2010ia,DelDebbioL,Degrand:2010ba}. The salient feature to be elucidated is the occurrence of a 
zero temperature phase transition in QCD when the number of flavor grows large,
but still below the loss of asymptotic freedom\cite{miransky_conformal_1997}.
This  phase transition separating the hadronic phase of QCD from a symmetric, conformal phase is a predictions from different model and analytic studies, and still needs a quantitative, first principle assessment.
To briefly summarize the status of the field at the time of this meeting, 
there is little doubt that SU(3) theories with a number of flavors below or equal eight are in the ordinary, broken phase of QCD \cite{Appelquist:2007hu,Deuzeman:2008sc,Jin:2008rc}. QCD with twelve flavors appears to be a borderline case, 
either from an analytic and numerical  point of view: our results \cite{Deuzeman:2009mh} as well as results based on the analysis of the Schroedinger functional
\cite{Appelquist:2007hu} favor chiral symmetry restoration
at weak coupling, and indicate the
existence of an IRFP, others suggest chiral symmetry 
breaking\cite{Fodor:2009wk}. According to a MCRG analysis
\cite{Hasenfratz:2010fi}
 the data favor the existence of an infrared fixed point and conformal phase, 
though the results are also consistent with very slow walking. 

It might be worthwhile to recall that conformality implies chiral symmetry -- 
indeed the conformal window has been uncovered  by noticing that,
in the presence of an IRFP,  the running coupling  might not grow large enough 
to break chiral symmetry \cite{appelquist_1996}. Only at a later stage it was 
recognized that the chiral transition
at the critical number of flavor was of a peculiar nature - dubbed {\em conformal}
\cite{miransky_conformal_1997}.
Hence, chiral symmetry restoration in the asymptotically free region of QCD,
and conformality go hand-in-hand. 
The direct observation of the physics of a conformal transition and the
behaviour at IRFP are of course an important subject of investigation,
which we are not addressing here. 

\section{The simulations, their systematics and the bulk transition}

We have simulated an  $SU(3)$ gauge theory with twelve flavors of
staggered fermions in the  fundamental representation.  We used
 a tree level  Symanzik improved gauge action to suppress lattice
artifacts,   and  Kogut-Susskind   (staggered)  fermions   with   the  Naik
improvement scheme.  High statistics runs were performed at fixed bare quark mass $am= 0.05$ over an extended range of bare lattice couplings, on $16^3\times8$ and  $16^4$ lattices. The results for the 
chiral condensate \cite{Deuzeman:2009mh}
suggest a bulk phase transition at $g_L = 1.35(3)$ .  

At two selected couplings, $6/g_L^2 = 3.9$ and $6/g_L^2 = 4.0$,
on the weak coupling side of the bulk transition, 
we have then performed runs on lattices $20^3 \times 32$, 
$24^4$, $32^4$ and five masses $am= 0.025, 0.04, 0.05, 0.06, 0.07$.
The thermalization of all runs was extensively verified by monitoring the stability of
averages and uncertainties as a function of the discarded number of sweeps, 
and bin size. In addition, we have verified the decorrelation
from initial conditions by performing simulations with ordered
and random starts for a few selected couplings and masses.

\begin{figure}
\vspace*{0.5 truecm}
\begin{center}
\includegraphics[width=9 truecm]{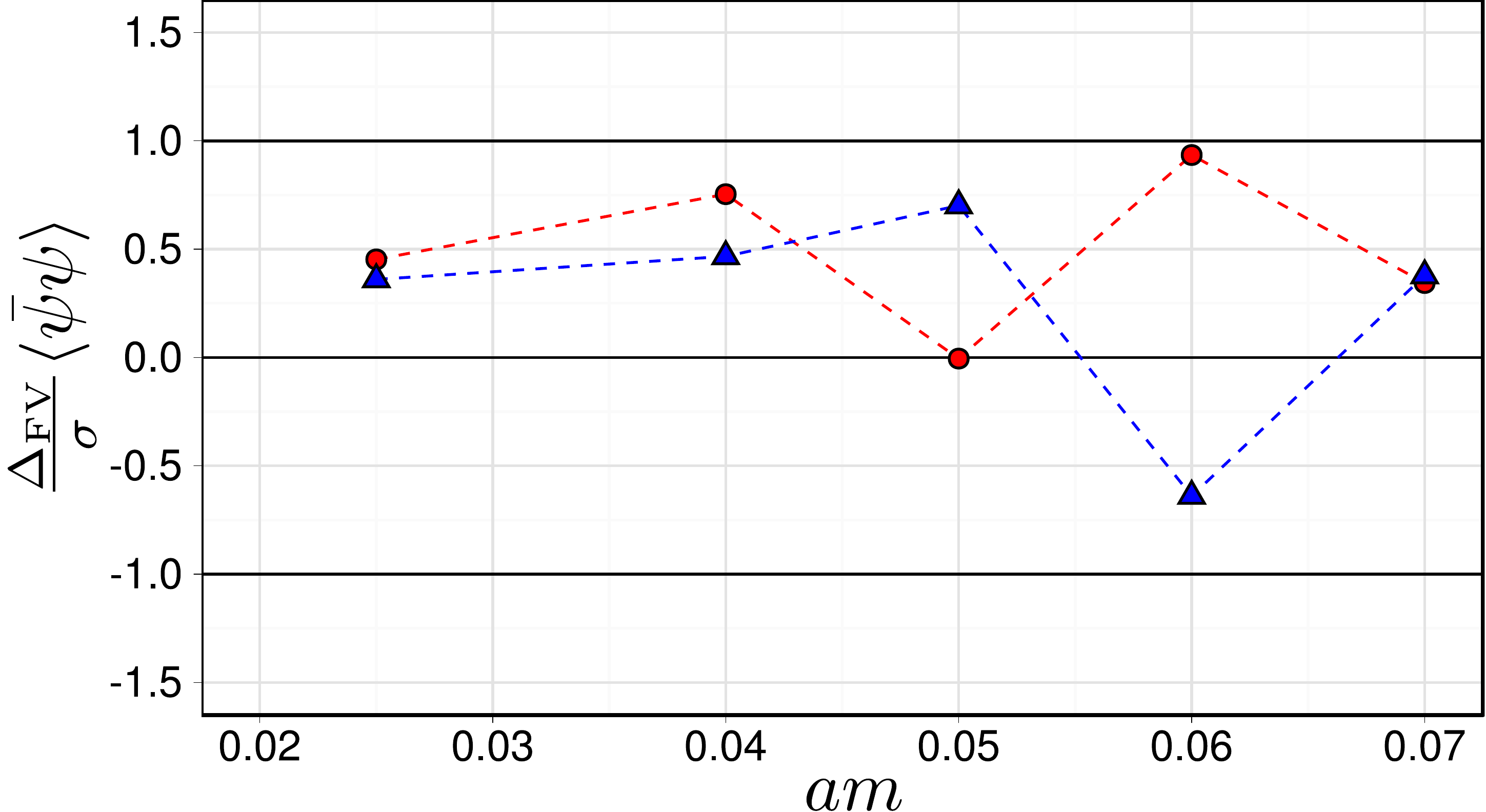}
\end{center}
\caption{\label{fig:fv_effect}  {(color   online)  Observed  finite  volume
effects   in   the  chiral   condensate,   displayed   as  the   difference
$\Delta_\mathrm{FV}$ between the measurements  at the two largest available
volumes ($24^3 \times  24$ and $32^3 \times 32$ for  the lowest mass, $20^3
\times 32$  and $24^3  \times 24$  for the other  masses) divided  by their
combined standard  deviation $\sigma$. Blue triangles  indicate results for
$6/g_L^2=3.9$, red circles those for  $6/g_L^2=4.0$.}}
\end{figure}

Measurements of the chiral  condensate 
were performed on  three different volumes for each mass,  up to $32^4$ for
the smallest  masses.  The data  set used
for the extrapolation to the chiral limit  consists of the measurements
at  lattice  volumes $24^4$,  which can  be
considered  as  infinite volume  measurements  within  their errors, 
see Fig.~\ref{fig:fv_effect}.

\section{The analysis of the chiral condensate}
Having observed a behaviour strongly suggestive of a bulk phase transition,
we examine in detail the chiral condensate, and the spectrum (next Section)
for two selected coupling on the weak coupling side of the transition itself.
Why staying close to the bulk transition? For one thing, the finite volume effects (which are expected to be large in a symmetric phase), 
should anyway be smaller
at larger coupling. Second, close to the bulk transition the system should
be QED-like and we can use previous expertise coming from the lattice analysis 
of the QED strong coupling transition. On a more technical note, 
we underscore that staggered fermions have  a remnant of exact chiral symmetry
which allows a  precise definition of the chiral order parameter -- 
the condensate $\langle\bar\psi\psi\rangle$ -- also on a coarse lattice,
with a genuine Goldstone modes and mass gap when the symmetry is broken.
\subsection{Fits motivated by a possible Goldstone phase}
\label{sec:goldstone}
The functional forms discussed here would be appropriate if the bulk behavior were not
to be associated to a true chiral transition. For instance, it might just 
be due to a generic rapid crossover, or to a genuinely lattice transition between two 
phases with different ordering. In this case the range of couplings 
between $6/g_L^2 = 3.9$ and $6/g_L^2 = 4.0$  would still belong 
to the phase with broken chiral symmetry. 
We have thus considered the following functional form:
\begin{equation}    
\langle\bar \psi \psi\rangle  = Am + Bm \log (m) + \langle\bar \psi \psi\rangle_0\, ,
\label{eq:goldstone}
\end{equation}
where the parameters were all left free, giving 
fits with two degrees of freedom, or in turn constrained to
zero. The logarithmic mass dependence is typical of a chirally broken phase for a 
QCD-like theory in four dimensions at zero temperature.

The results of the fits to Eq.~\ref{eq:goldstone}
are summarized in Table~\ref{tab:goldstonefit}. 
The linear fits  
produce an intercept different from zero, 
but are highly disfavored by their large $\chi^2$. 
The inclusion of the term $m \log (m)$ considerably improves
the quality of the fits. Those with free intercept  $\langle\bar \psi \psi\rangle_0$
gave an extrapolated value consistent with zero, and in agreement with the fit obtained 
by constraining $\langle\bar \psi \psi\rangle_0 =0$. Both fits are satisfactory,
and imply that 
the chiral condensate in the chiral limit is zero within errors. 
In conclusion, a conventional picture of the Goldstone phase
seems not to be supported by our data.
\begin{table}
\begin{center}
\begin{tabular}{|c|c|c|c|c|}
$6/g_L^2 $ & A & B &  $\langle\bar \psi \psi\rangle_0$ & $\sqrt{\chi^2\,\mathrm{dof}}$  \\
\hline
 3.9    & 2.70(3) & -0.103(13) & 0.00013(54) & 0.68  \\
        & 3.12(3)   &  0 (F) & 0.0043(3)   & 3.12  \\  
        & 2.682(5) & -0.107(2) & 0 (F)   & 0.56 \\
\hline
 4.0    & 2.48(2) & -0.120(10) & -0.00091(42) & 0.51  \\
        & 2.73(1) & 0 (F)  & 0.0041(5) & 3.74  \\
        & 2.519(8) & -0.099(3) & 0 (F) & 0.56  \\
\end{tabular}
\end{center}
\caption{\label{tab:goldstonefit}
Fits to $\langle\bar \psi \psi\rangle = A m + B m \log m + \langle\bar \psi \psi\rangle_0 $}
\end{table}

\subsection{Fits with an anomalous dimension}

We considered the functional form 
\begin{equation}
\langle\bar \psi \psi\rangle = A m ^{1/\delta} + Bm + \langle\bar \psi \psi\rangle_0\, ,
\label{eq:anomalous}
\end{equation}
 containing an anomalous dimension, whose effect is parameterized by the exponent $\delta$.
Since the fits described in section  \ref{sec:goldstone} already 
suggest that a curvature in the behavior of the chiral condensate as a
function of the mass  is mandatory, we started by setting the linear term  
to zero. We note that analogous fits were used in the past
 to analyze QED in its symmetric phase, close to the strong
coupling transition in Ref.~\cite{Kocic:1990fq},  
even if a more satisfactory account of the
data requires the consideration of the magnetic equation of state,
which is going to be discussed in the next section. 
All fits to Eq.~(\ref{eq:anomalous}) with $B=0$ are satisfactory, with a chiral condensate
compatible with zero in the chiral limit. This was checked, as before,
by comparing fits with free intercept, and fits with 
$\langle\bar \psi \psi\rangle_0 = 0$. 

One might still suspect that a fit combining a power-law term and a linear
term, with a non zero intercept might still  
accommodate the data, hence indicating chiral symmetry breaking.
For completeness we have performed fits to Eq.~(\ref{eq:anomalous}) with the inclusion of 
a linear term. As expected from the near degeneracy between a power law with 
$1/\delta \approx 1$ and a linear term, the uncertainties coming from a Marquardt-Levenberg 
minimization of $\chi^2$ are huge.

To acquire a feeling about the possible relevance of
a linear term, we have also performed a sequence of fits, constraining the
exponent to several values in the acceptable range given by
the fit errors.
It appears that the coefficient of the linear term  smoothly
changes from positive to negative, while the intercept - the chiral
condensate in the chiral limit - remains consistent with zero throughout
at $6/g_L^2=3.9$, and becomes slightly negative at $6/g_L^2=4.0$.
We thus again conclude that our data point at exact chiral symmetry.

In conclusion,  regardless the presence of 
any additional linear term, or any analytic term in Eq.~(\ref{eq:anomalous})
the resulting extrapolated condensate remains compatible with zero. The interested reader is referred
again to our extended publication Ref.\cite{Deuzeman:2009mh} for details. 

\subsection{Fits motivated by the Magnetic Equation of State}
\label{sec:EOS}
Finally we considered fits motivated by the magnetic equation of state,
see e.g. Ref.~\cite{Kocic:1990fq} and references therein. 
The following equation is a satisfactory parameterization
\begin{equation}
m = A \langle\bar \psi \psi\rangle + B \langle\bar \psi \psi\rangle^{\delta}\, ,
\label{eq:magneticeos}
\end{equation} 
which would of course coincide with the simple power law when A=0.
The coefficient of the linear term $A$ should vanish at a critical
point, with $A \propto (\beta - \beta_c)$. This of course explains the
smallness of A close to the transition, while
$\delta$ is the conventional magnetic exponent.
The linear term in the condensate is implied by chiral symmetry, and guarantees that 
the ratio 
\begin{equation}
\lim_{m \to 0} R_\pi 
= \frac{\partial\langle\bar\psi\psi\rangle /\partial m}{\langle\bar \psi \psi\rangle/m}
= 1
\end{equation}
approaches unity in the chiral limit and in the chirally symmetric phase.
We can view Eq.~(\ref{eq:magneticeos}) as a model for a theory
with anomalous dimensions, which incorporates the correct chiral limit.
Note that the linear term
of Eq.~(\ref{eq:magneticeos}) is of different origin than the one considered
in Eq.~(\ref{eq:anomalous}). The latter describes violations of
scaling and it is increasingly relevant at larger masses. In  Eq.~(\ref{eq:magneticeos}) 
instead, it is dominating at very small masses, away from the critical
point.

Results for this case are given in Table \ref{tab:eosfit}.
The fit $m = m(\langle\bar \psi \psi\rangle)$  
was performed with a least squares algorithm. Note that, as expected, the
significance of the linear term is very low, closer to the bulk
transition, and slightly larger by moving away from it. 
Indeed,  the agreeement between the numerical solutions of the
equation $m( \langle\bar \psi \psi\rangle ) = m_{sim}$, with $m_{sim}$ the simulation 
masses, with the simulation results for the condensate is very good.

All fits clearly favor a positive value for the coefficient
of the linear term, as it should be in the chirally symmetric phase,
and within the large errors the results for the exponent are compatible
with the ones coming from the genuine power law fits. We conclude again
 in favor of chiral symmetry restoration. 
\begin{table}
\begin{center}
\begin{tabular}{|c|c|c|c|}
$6/g_L^2 $ & A & B &  $\delta$  \\
\hline
 3.9    &  0.1(9) & 0.3(9)  & 1.1(2) \\
\hline
 4.0    &  0.3(1) & 0.077(9) & 1.3(1) \\
\end{tabular}
\end{center}
\caption{\label{tab:eosfit}
Fits to $m = A\langle\bar \psi \psi\rangle + B\langle\bar \psi \psi\rangle^\delta $}
\end{table}

\section{Spectrum analysis}
\label{sec:spectrum}

\begin{figure}[t]
\hspace*{-0.5 truecm}
\subfigure[Interpolations    of   meson   masses.]{\label{fig:Spectral_fit}
\includegraphics[width=8.0truecm]{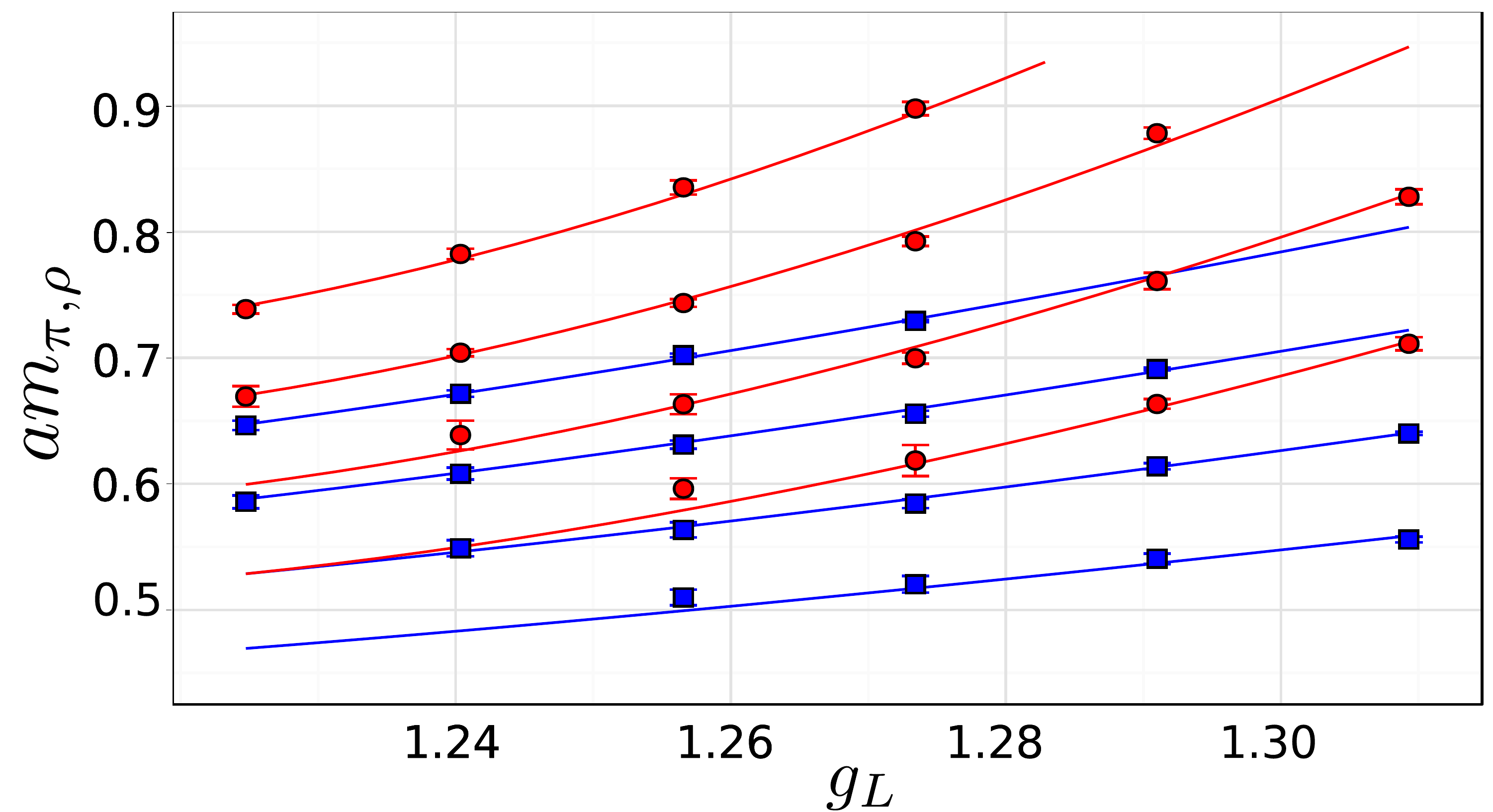}}   \subfigure[$\pi$  to
$\rho$    mass   ratio    with   fits    superimposed   ]{\label{fig:Ratio}
\includegraphics[width=8.0truecm]{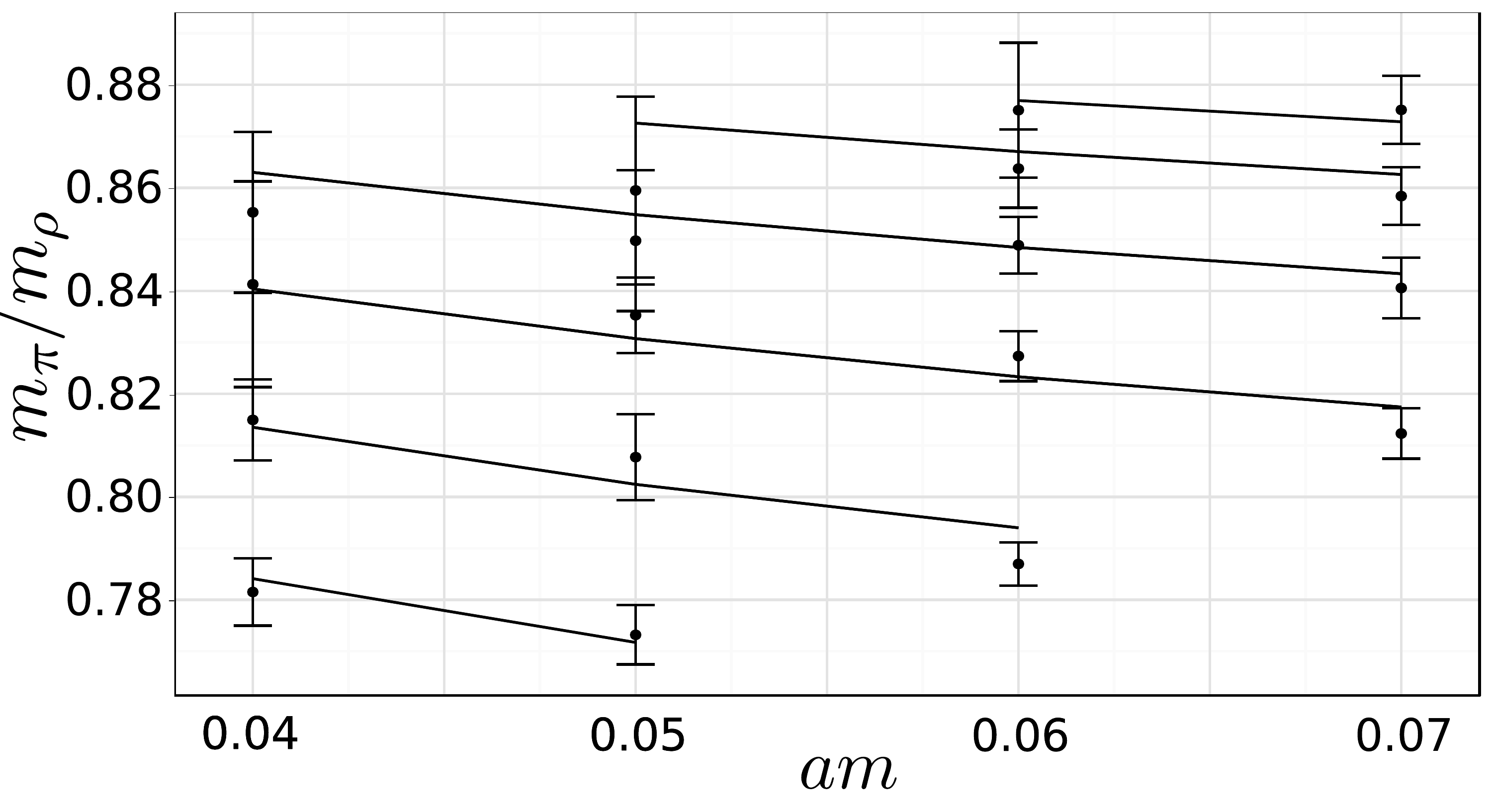}}
\caption{(color  online) (a)  Measurements of  the pseudoscalar  (blue) and
vector (red) masses  versus lattice coupling at several  values of the bare
quark  mass, from  bottom  to top  $am=0.04$,  0.05, 0.06  and 0.07,
with fits superimposed. (b)  The  measured $\pi$  to
$\rho$ mass  ratio as a function  of the bare mass  and decreasing coupling
$g_L$,  bottom to  top $6/g_L^2=3.5$  to  $4$. The  superimposed lines  are
ratios of the best fits.}
\end{figure}

We have fitted both the pion and the
rho mass to a power law
\begin{equation}
m_{\pi,\rho} = A_{\pi,\rho} m ^{\epsilon_{\pi,\rho}}
\end{equation}
with the results
$A_\pi = 3.41(21)$, $\epsilon_\pi = 0.61(2)$,  
$A_\rho = 4.47(61)$, $\epsilon_\rho = 0.66(5)$  at $6/g_L^2 = 3.9$, and 
$A_\pi = 3.41(21)$, $\epsilon_\pi = 0.61(2)$,
$A_\rho =  4.29(11)$, $\epsilon_\rho  = 0.66(1)$ at $6/g_L^2 =  4.0$. 
From these
fits -- also plotted in Fig.~\ref{fig:Spectral_fit} --
we conclude that  the mass dependence  of the vector and
pseudoscalar mesons is well fitted by a power-law.  Second, it is 
also relevant
that  the exponents are  not unity  and $\epsilon_\pi  \neq 1/2$.   The  latter result
immediately tells  that the pion  seen here is  not a Goldstone boson  of a
broken chiral symmetry.  

By looking at the behavior
of the mass ratio we can further inspect the status of chiral symmetry. 
Fig.~\ref{fig:Ratio} shows  the ratios of measured  pseudoscalar and vector
masses, for a fixed  coupling and as a function of the  bare quark mass. We
have superimposed  the ratios  of the best  fits to  the raw mass  data.   
It is immediately clear that the
ratio increases as  the quark mass approaches zero,  a behavior opposite to
what is expected for a Goldstone pion (see  Ref.~\cite{Kocic:1990fq} for
more exteded discussions) .  

Note that the spectrum results can be used  to  determine the lines of 
``constant physics'' in
the two dimensional parameter space $g_L$  and $am$, the bare quark mass of
degenerate fermions, following the same strategy which was successful 
for $N_f = 16$~\cite{damgaard_lattice_1997}. 

\section{\label{sec:conclusion}Summary and Outlook}

After observing a bulk transition for $N_f = 12$,  we are continuing our study along two lines:  we explored the nature of the weak coupling phase, which is the subject of this note, and we are investigating in more details the nature
of this bulk transition, see the contribution by Elisabetta Pallante\cite{PallanteE}. 

The results presented here favor chiral symmetry restoration in the range
of couplings which we have explored. Barring unexpected re-entrant transitions
at even weaker coupling (a preliminary $\beta$ scan shows no 
sign of such phase transitions) this should indicate chiral symmetry also in
the continuum limit, compatible with the existence of an IRFP. 

Can we challenge this result?  It is  clear that additional data  at even 
lighter masses or extremely larger volumes  might in principle  leave room 
for different conclusions. 
In practice,  it seems difficult to reverse the trend of the  mass ratio. 

If the $N_f=12$ theory is very close to the critical number of flavor
it is indeed rather natural to have ambiguous results from different methods
and setups.  The analysis of the phase diagram in the Temperature-$N_f$ space \cite{Braun:2009ns} might help understanding why: 
if the critical exponent of the running gauge coupling at the IRFP
is less then one,  the critical temperature becomes exponentially small 
when approaching $N_f^c$  : a cold quark gluon plasma will 
then become arbitrarily close to the zero temperature conformal window,
and the hadronic phase will only appear for a very large $N_t$. 

A conservative conclusion - encompassing our results as well as those from other groups - is then that the critical number of flavor is very close to twelve, and a speculation is that the critical exponent of the running gauge coupling at the Banks-Zacs fixed point is less than one.

\bibliographystyle{JHEP} 
\bibliography{lat2010}

\end{document}